\documentclass{elsarticle}
\usepackage{graphicx}
\usepackage{amsmath,amssymb,yhmath,makeidx,mathdots,setspace,framed,young,lscape}
\usepackage[enableskew]{youngtab}  
\usepackage[usenames]{color}

\newcommand{\trnm}[1]{\mbox{#1}}
\newcommand{\ncd}{\newcommand}
\ncd{\END}{\trnm{End}}
\ncd{\nn}{\nonumber}
\ncd{\ba}{\begin{array}}
\ncd{\ea}{\end{array}}
\ncd{\be}{\begin{equation}}
\ncd{\ee}{\end{equation}}
\ncd{\bea}{\begin{eqnarray}}
\ncd{\eea}{\end{eqnarray}}
\ncd{\ga}{\alpha}
\ncd{\gb}{\beta}
\ncd{\GG}{\Gamma}
\ncd{\gl}{\lambda}
\ncd{\GL}{\Lambda}
\ncd{\grg}{\gamma}
\ncd{\go}{\omega}
\ncd{\GO}{\Omega}
\ncd{\gr}{\rho}
\ncd{\nl}{\nabla}
\ncd{\gre}{\epsilon}
\ncd{\td}{\widetilde}
\ncd{\st}{\longrightarrow}
\ncd{\stk}{\rightarrow}
\ncd{\na}{\longmapsto}
\ncd{\fa}{\forall}
\ncd{\cc}{\circ}
\ncd{\za}{\subset}
\ncd{\wtw}{\Longleftrightarrow}
\ncd{\pcz}{\partial}
\ncd{\gz}{\zeta}
\ncd{\gt}{\theta}
\ncd{\gvt}{\vartheta}
\ncd{\gs}{\sigma}
\ncd{\GS}{\Sigma}
\ncd{\cl}{{\cal L}}
\ncd{\gd}{\delta}
\ncd{\GD}{\Delta}
\ncd{\ld}{\trnm{Ldiff}}
\ncd{\cj}{{\cal J}}
\ncd{\ce}{{\cal E}}
\ncd{\cf}{{\cal F}}
\ncd{\tr}{{\trnm Tr}}
\ncd{\bgw}{{\bigwedge}}
\ncd{\im}{\trnm{im}}
\ncd{\szesc}{{\times _6}}
\ncd{\siedem}{{\times _7}}
\ncd{\zlap}{\frac{\pcz ^2}{ \pcz z \pcz z^*}}
\ncd{\ztap}{\frac{\pcz }{ \pcz t}}
\ncd{\pczz}{\frac{\pcz}{\pcz z}}
\ncd{\pczsp}{\frac{\pcz}{\pcz z^*}}
\ncd{\zlapkp}{\frac{\pcz ^2}{ \pcz z_{kp} \pcz {z_{kp}^*}} }
\ncd{\bra}{\langle}
\ncd{\ket}{\rangle}
\ncd{\mnp}{M}
\ncd{\gk}{\overline}
\ncd{\przh}{\mathcal{H}}
\ncd{\przhr}{\mathcal{H}^{(r)}}
\ncd{\cq}{\mathbb{Q}}
\ncd{\cqo}{\mathbb{Q}(\omega)}
\ncd{\autq}{\mathrm{Aut}\,\mathbb{Q}(\omega)}
\ncd{\thl}{\Theta_l}
\ncd{\mpdwa}{\mp 2}
\ncd{\pmdwa}{\mp 2}
\ncd{\pmjed}{\mp 1}
\ncd{\mdwa}{-2}
\ncd{\Hom}{\trnm{Hom}}
\ncd{\bft}{\mathbf{t}}
\ncd{\bfj}{\mathbf{j}}
\ncd{\httz}{\mathcal{H}^{33}_0}

\begin{document}

\begin{frontmatter}

\title{Internal parity symmetry and degeneracy of Bethe Ansatz strings in the isotropic heptagonal magnetic ring}
\author[jm]{J. Milewski}
\ead{jsmilew@wp.pl}
\author[bl]{B. Lulek}
\ead{barlulek@amu.edu.pl}
\author[tlp,bl]{T. Lulek}
\ead{tadlulek@prz.edu.pl}
\author[ml]{M. \L{}abuz\corref{mlab}}
\ead{labuz@univ.rzeszow.pl}
\author[rs]{R. Stagraczy\'nski}
\ead{rstag@prz.edu.pl}
\cortext[mlab]{Corresponding author: Tel: +48 17 872 11 06, Fax: +48 17 872 12 83}
\address[jm]{Institute of Mathematics, Pozna\'n University of Technology, \\Piotrowo 3A, 60-965 Pozna\'n, Poland}
\address[bl]{East European State Higher School, ul. Tymona Terleckiego 6, 37-700 Przemy\'sl, Poland}
\address[tlp]{Faculty of Physics, Adam Mickiewicz University, \\Umultowska 85, 61-614 Pozna\'n, Poland}
\address[ml]{University of Rzeszow, Institute of Physics, Rejtana 16a, 35-959 Rzesz\'ow, Poland}
\address[rs]{Rzeszow University of Technology, The Faculty of Mathematics and Applied Physics, Powsta\'nc\'ow Warszawy 6, 35-959 Rzesz\'ow, Poland}

\begin{abstract}
The exact Bethe eigenfunctions for the heptagonal ring within the isotropic XXX model exhibit a doubly degenerated energy level in the three-deviation sector at the centre of the Brillouin zone. We demonstrate an explicit construction of these eigenfunctions by use of algebraic Bethe Ansatz, and point out a relation of degeneracy to parity conservation, applied to the configuration of strings for these eigenfunctions. Namely, the internal structure of the eigenfunctions (the 2-string and the 1-string, with opposite quasimomenta) admits generation of two mutually orthogonal eigenfunctions due to the fact that the strings which differ by their length are distinguishable objects.
\end{abstract}


\end{frontmatter}

\section{Introduction}
The famous Bethe Ansatz (BA) solution \cite{bethe} of the eigenproblem of the Heisenberg Hamiltonian of a finite magnetic ring of $N$ nodes with the spin $1/2$ and isotropic nearest-neighbour interaction (the XXX model) is formulated in terms of the hypothesis of strings \cite{bethe} - \cite{yangone}. This hypothesis was taken to hold in the limit $N\to \infty$, but has been a posteriori found to be essentially correct, with some deformations, in the most of cases for finite $N$ \cite{yangone} - \cite{hage}. Within this picture, an exact BA eigenstate of the highest weight is specified by the so called rigged string configuration $\nu\mathcal{L}$ \cite{kkr} - \cite{llls}. In more detail, a highest weight state is such that the number $r$ of reversed spins (or Bethe pseudoparticles) is equal to $N/2-S$, where $S$ is the quantum number of the total spin of the magnet. Then, $\nu$ is a partition of $r$, referred to as {\em the string configuration}: each row of $\nu$ is a string, whose {\em length} is given by the number of boxes in this row. Finally, each string is equipped with its own quasimomentum, referred to as {\em the rigging}, and $\mathcal{L}$ denotes the collection of all riggings for a given BA eigenstate. Therefore, each exact BA eigenstate has the interpretation of a collection of strings of various lengths, represented by the partition $\nu \vdash r$, and each string has its own exact quantum number, i.e. rigging by a quasimomentum. 

Such a string interpretation of BA eigenstates was recently supported by an explicit calculation of exact values of related spectral parameters specyfying the eigenstate for the case of magnetic pentagon \cite{mbll}. It was also shown that Galois groups of the associated number field extensions of the prime field $\mathbb{Q}$ of rationals (responsible for the eigenproblem of the Heisenberg Hamiltonian in the initial basis of all magnetic configurations) by the exact values of spectral parameters acquire a natural interpretation of permutations on the set of $r$ boxes of string configurations $\nu$ ($r=2$ for the case of pentagon, $N=5$). In particular, some Galois symmetries are responsible for transmutation of bound $(\nu=\{2\})$ and scattered  $(\nu=\{1^2\})$ two-magnon states. 

In the present paper we aim to extend further such an interpretation of BA eigenstates as literal realizations of rigged string configurations, to a specific case of two-dimensional subspace of the state space of heptagon ($N=7$), with degenerated energy ($E=-5$), quasimomentum ($k=0$, the centre of the Brillouin zone), and the total spin ($S=1/2$, and thus the three-magnon sector $r=3$). This space realizes a particular case of an ''arithmetic qubit'' in the terminology of Ref. \cite{mblls}. We first perform the exact diagonalization of the Heisenberg Hamiltonian for the three-magnon sector using the standard basis of wavelets \cite{llwj}, next determine the corresponding spectral parameters within so called ''inverse BA'' \cite{mblls} - \cite{llwj}, then apply algebraic BA \cite{faddeevrus} - \cite{fadd} for an explicit construction of desired eigenstates as symmetric functions of spectral parameters, and last, discuss the properties of obtained exact forms of density matrices, in particular their behaviour under the parity operation.

\section{The eigenproblem of the Heisenberg Hamiltonian in the three-magnon sector at the centre of the Brillouin zone}

a) The Brillouin zone of the heptagon.\\

Let $\Gamma_k$, specified by 
\be
\Gamma_k(j)=\omega^{kj}, \quad j\in \tilde{7}=\{1,\ldots,7\},
\ee
where $\omega=\exp(2\pi i/7)$ is the first primitive 7-th root of unity, be an irreducible representation of the cyclic group $C_7$ - the translational symmetry group of the heptagon, and let
\be
B=\{k=0,\pm 1,\pm 2, \pm 3\}
\ee
be the set of labels of all such representations. Clearly, $B$ is the dual group to $C_7$, or, in physical terms, the Brillouin zone for the heptagon, with elements $k\in B$ recognized as quasimomenta - exact quantum numbers which reflect the translational symmetry of the model.\\

b) Some invariant subspaces.\\

Let $r\in \{0,1,2,3\}$ be the number of spin deviations, i.e. Bethe pseudoparticles (we consider only the states ''below equator''), $r'\in\{0,\ldots,r\}$ be the number of those Bethe pseudoparticles which are coupled into strings (so that $r=r'$ corresponds to highest weight states), and

\be
\label{arrows}
\begin{array}{ccccccc}
&&\mathcal{H}^{rr'}&&&&\\
&\nearrow&&\searrow&&&\\
\mathcal{H}^{rr'}_k&&&&\mathcal{H}^r&\to&\mathcal{H}\\
&\searrow&&\nearrow&&&\\
&&\mathcal{H}^{r}_k&&&&\\
\end{array}
\ee

\noindent be a scheme which displays some relevant subspaces of the state space $\mathcal{H}$ for the heptagon, with specified quantum numbers $r$, $r'$ and $k$, or, equivalently, the total magnetization $M=7/2-r$, the total spin \linebreak $S=7/2-r'$, and the total quasimomentum $k\in B$. An arrow in (\ref{arrows}) indicates the relation subspace $\to$ space. In the present paper, we are interested in the scheme

\be
\label{arrows2}
\begin{array}{ccccccc}
&&\mathcal{H}^{33}&&&&\\
&\nearrow&&\searrow&&&\\
\mathcal{H}^{33}_0&&&&\mathcal{H}^3&\to&\mathcal{H}\, ,\\
&\searrow&&\nearrow&&&\\
&&\mathcal{H}^{3}_0&&&&\\
\end{array}
\ee

\noindent with the dimensions given by


\be
\begin{array}{c}
\mathrm{dim\,}\mathcal{H}^{33}_0=2 \quad \mathrm{dim\,}\mathcal{H}^{3}_0=5 \quad \mathrm{dim\,}\mathcal{H}^{33}=\left({7 \atop 3}\right)-\left({7 \atop 2}\right)=14 \\[+0.15cm]
\mathrm{dim\,}\mathcal{H}^{3}=\left({7 \atop 3}\right)=35 \quad  \mathrm{dim\,}\mathcal{H}=2^7=128\, .
\end{array}
\ee

Our space of interest is $\mathcal{H}^{33}_0$, a two-dimensional subspace with degenerated values of energy ($E=-5$), total spin ($S=1/2$), and quasimomentum ($k=0$). In the terminology of Ref. \cite{mblls}, $\mathcal{H}^{33}_0$ is an example of an ''arithmetic qubit'', which can be implemented on the state space $\mathcal{H}$ of the heptagon.\\

c) The basis of wavelets.

The initial (calculational) basis for the arithmetic qubit $\mathcal{H}^{33}_0$ can be specified in terms of embedding  $\mathcal{H}^{33}_0\to  \mathcal{H}^{3}_0$ in the three-magnon sector corresponding to the centre $k=0$ of the Brillouin zone. We choose the basis of wavelets in $\mathcal{H}^{3}_0$, in accordance with Ref. \cite{llwj}. Essentially, this is the basis obtained from the set of all magnetic configurations with 3 spin deviations, that is the classical configuration space for the system of $r=3$ Bethe pseudoparticles on the heptagon $\tilde{7}$, reduced by the natural action of the translation group $C_7$. This action generates $5=35/7$ regular $C_7$-orbits, specified by relative positions $\bft=(t_1,t_2,t_3)$ of the system of Bethe pseudoparticles on the ring $\tilde{7}$. Clearly, a triad $\bft$ is subject to the following constraints:
\begin{enumerate}
\item $t_{\alpha}=(j_{(\alpha+1)\mathrm{\, mod \,}3}-j_{\alpha})\mathrm{\, mod \,}7$ denotes the distance between consecutive \linebreak ($(\alpha+1)\mathrm{\, mod \,}3$ and $\alpha$) Bethe pseudoparticles on $\tilde{7}$, such that
\be
\sum_{\alpha\in\tilde{3}}t_{\alpha}=7.
\ee
\item Those triads which differ by a cyclic permutation, i.e. $(t_1,t_2,t_3)$, $(t_2,t_3,t_1)$, and $(t_3,t_1,t_2)$, give rise to the same $C_7$ orbit, and are thus equivalent; we choose the triad which is lexically the first.
\end{enumerate}

The basis of wavelets for the space $\mathcal{H}^{3}_0$ is presented in Fig. 1.

\begin{figure}[ht]
\begin{center}
\includegraphics*[width=0.35\linewidth]{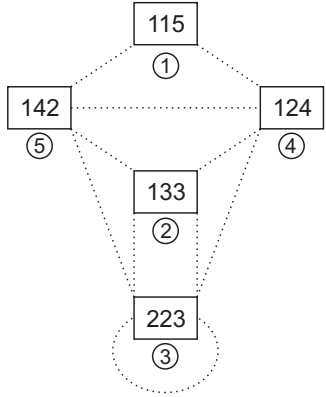}
\label{fig_basisk}
\end{center}
\caption{The basis of wavelets for the space $\mathcal{H}^3_0$. Each rectangle $(t_1,t_2,t_3)$ labels a wavelet. The encircled number below the rectangle defines labeling of rows and columns of matrices in the main text. Each dashed line indicates an interaction channel for the Heisenberg Hamiltonian.}
\end{figure}

Within the conventions of this figure, the projector $P^{33}_0$ from the space $\mathcal{H}^3_0$ onto the qubit $\mathcal{H}^{33}_0$, such that
\be
\label{eqeight}
P^{33}_0\, \mathcal{H}^3_0=\mathcal{H}^{33}_0, \quad (P^{33}_0)^2=P^{33}_0,
\ee
is readily obtained as
\be
\label{eqeightt}
P^{33}_0=\frac{1}{15}
\left(
\begin{array}{cccccc}
2&2&2&\shortmid&-3&-3\\
2&2&2&\shortmid&-3&-3\\
2&2&2&\shortmid&-3&-3\\
\mbox{--}&\mbox{--}&\mbox{--}&\mbox{+}&\mbox{--}&\mbox{--}\\
-3&-3&-3&\shortmid&12&-3\\
-3&-3&-3&\shortmid&-3&12\\
\end{array}
\right)
\ee
This simple result suggest us to decompose the set $V$ of all 5 wavelets of Fig. 1 into subsets 
\be
\label{eqnine}
\begin{array}{c}
V_1=\{1,2,3\}\equiv\{(1,1,5),(1,3,3),(2,2,3)\}, \\[+0.1cm]
V_2=\{4,5\}\equiv\{(1,2,4),(1,4,2)\},
\end{array}
\ee
so that each element of a subset enters the projector $P^{33}_0$ on equal footing: the same diagonal elements, i.e. probabilities ($2/15$ and $12/15$ for $V_1$ and $V_2$, resp.), the same inner hybridizations ($2/15$ and $-3/15$ for $V_1$ and $V_2$, resp.), and a single outer hybridization $-3/15$. Moreover, these subsets exhibit distinct behaviour under the parity operator on the heptagon $\tilde{7}$, i.e. the reflection in the node $j=7\equiv 0\mathrm{\, mod\, }7$, given by the permutation
\be
\label{eqten}
\pi=
\left(
\begin{array}{ccccccc}
1&2&3&4&5&6&7\\
6&5&4&3&2&1&7
\end{array}
\right) .
\ee
The related parity operator in the space $\mathcal{H}^3_0$ (denoted for simplicity also by $\pi$) is represented in the basis $V$ of wavelets as
\be
\label{eqeleven}
\pi=
\left(
\begin{array}{cccccc}
1&&&\shortmid&&\\
&1&&\shortmid&&\\
&&1&\shortmid&&\\
\mbox{--}&\mbox{--}&\mbox{--}&\mbox{+}&\mbox{--}&\mbox{--}\\
&&&\shortmid&&1\\
&&&\shortmid&1&
\end{array}
\right),
\ee
so that each element of $V_1$ is a $\pi$-invariant, whereas $V_2$ is a regular orbit of the two-element reflection group. \\

d) The eigenproblem of the Hamiltonian.\\

The Heisenberg Hamiltonian in the basis of magnetic configurations is a {\em local} operator. In fact, one has
\be
\label{eqtw}
\hat{H}|\bfj\rangle=\sum_{\bfj'}(|\bfj'\rangle-|\bfj\rangle),
\ee
where $|\bfj\rangle=|j_1,\ldots,j_r\rangle$ is a state with specified positions $j_{\alpha}\in\tilde{7}$, $\alpha\in\tilde{r}$, of Bethe pseudoparticles, and the sum in rhs of Eq. (\ref{eqtw}) runs over such $\bfj'$ which are the nearest neighbours of  $\bfj$, i.e. they differ from  $\bfj$ at only one argument, say $j_{\alpha}$, and $j_{\alpha}'=(j_{\alpha}\pm 1)\mathrm{\, mod\,}7$. It means that each non-diagonal matrix element $\langle\bfj'|\hat{H}|\bfj\rangle=1$, referred to an interaction channel, is accompanied by the diagonal contribution $\langle\bfj|\hat{H}|\bfj\rangle=-1$. In the case of the space $\mathcal{H}^3_0$, it results in the Hamiltonian matrix
\be
\label{eqth}
H=
\left(
\begin{array}{cccccc}
-2&0&0&\shortmid&1&1\\
0&-4&2&\shortmid&1&1\\
0&2&-4&\shortmid&1&1\\
\mbox{--}&\mbox{--}&\mbox{--}&\mbox{+}&\mbox{--}&\mbox{--}\\
1&1&1&\shortmid&-4&1\\
1&1&1&\shortmid&1&-4\\
\end{array}
\right).
\ee
This matrix can be readily interpreted using Fig. 1
, with all interaction channels indicated by dashed lines, joining appropriate wavelets $\bft$ and $\bft'$. It is worth to observe that (i) each element between $\bft$ and $\bft'$ yields the contribution $+1$ to the matrix element $\langle\bft'|\hat{H}|\bft\rangle$, and thus nondiagonal elements in (\ref{eqth}) take on the values $0$, $1$, or $2$; (ii) the number of channels outgoing a wavelet $\bft$ is equal to the doubled number of islands of adjacent Bethe pseudoparticles on the heptagon $\tilde{7}$; it is equal to $2$, $4$, $6$, $4$, $4$ for $\bft$ labeled in Fig. 1 
by $1$, $2$, $3$, $4$, $5$, respectively (it follows from the fact that internal Bethe pseudoparticles in an island are kinematically frozen in the mechanism ruled by Eq. (\ref{eqtw})); the diagonal elements $\langle\bft|\hat{H}|\bft\rangle$ are equal to minus the doubled number of outgoing lines for $\bft$ labeled by $1$, $2$, $4$, and $5$, since these wavelets do not have any internal channels, whereas the wavelet $\bft=(2,2,3)$ has a single internal channel (i.e. \linebreak $\bft=\bft'$, but $\bfj\neq\bfj'$, a term corresponding to hybridization between different magnetic configurations $\bfj\neq \bfj'$ within the same $C_7$-orbit $\bft$), which yields \linebreak $\langle 3|H|3\rangle=-2\cdot 3\, +\, 2\cdot 1=-4$.

The characteristic polynomial of the Hamiltonian (\ref{eqth}) reads
\be
w^H(x)=x(x+2)(x+6)(x+5)^2,
\ee
which yields the spectrum $\mathrm{spec\, }H=\left\{0,-2,-6,-5^{(2)}\right\}$. The doubly degenerated eigensubspace of $\mathcal{H}^3_0$ with $E=-5$ is readily identified with the qubit $\mathcal{H}^{33}_0$. Now our problem consists in finding the basis of BA eigenstates for this qubit. 


\section{Spectral parameters of Bethe eigenstates}
In most cases, an exact eigenstate of the Heisenberg Hamiltonian has \linebreak  a unique assignment of energy $E$, total spin $S$, magnetization $M$ and quasimomentum $k$, so that examination of Bethe string hypothesis may be performed by scanning appropriate spectral parameters on a single eigenstate. Here we consider an exceptional case in this respect since the degeneracy admits such eigenstates which cannot be presented in the form required by BA. It rises a natural question how to select states of the BA form within this qubit.

As mentioned in the Introduction, a Bethe eigenstate $|\nu\mathcal{L}\rangle$ is completely characterized by the string configuration $\nu\vdash r'$, and its rigging $\mathcal{L}$. Analytical form of such an eigenstate is described in terms of $r'$ spectral parameters $\lambda_{\alpha}$, $\alpha\in \tilde{r'}=\{1,\ldots,r'\}$, or, equivalently, portions of phase $a_{\alpha}$. The latter is defined by
\be
\label{eqf}
a=e^{ip}=\frac{\lambda+\frac{i}{2}}{\lambda-\frac{i}{2}},
\ee
where $p$ is known as the {\em pseudomomentum}. For $a\neq 1$, i.e. $\lambda\neq\pm\infty$, or $p\neq 0$, the inverse of Eq. (\ref{eqf}) (the Cayley transform) reads
\be
\label{eqjacobi}
\lambda=\frac{i}{2}\frac{a+1}{a-1}.
\ee
We proceed to derive, along the so called ''inverse BA'', the portions of phase, denoted by $a$, $b$, $c$, within the degenerated qubit $\mathcal{H}^{33}_0$. The conservation of quasimomentum for the centre of the Brillouin zone reads
\be
\label{eqs}
abc=1,
\ee
and the conservation of energy yields
\be
a+a^{-1}+b+b^{-1}+c+c^{-1}=E=-5,
\ee
whereas the Bethe equation reads 
\be
\label{eqn}
a^7=\frac{(ab-2a+1)(ac-2a+1)}{(ab-2b+1)(ac-2c+1)}.
\ee
Eqs. (\ref{eqs})-(\ref{eqn}) yield a single polynomial equation of one variable, say, $t$, in the form
\be
\label{eqtz}
f(t)\equiv t^6-t^5+5(t^4+t^3+t^2)-t+1=0.
\ee
The portions of phase corresponding to BA eigenstates should be therefore some roots of the polynom $f$, defined by Eq. (\ref{eqtz}). We analyze these roots in some detail, and demonstrate that they indeed determine exactly two eigenstates of the form required in BA within the qubit $\mathcal{H}^{33}_0$.

Coefficients of the polynom $f$ are invariant with respect to interchange \linebreak $t^i \, \leftrightarrows \, t^{6-i}$, so that the substitution 
\be
\label{eqto}
x=t+t^{-1}
\ee
yields the third order polynomial equation for the variable $x$, namely
\be
g(x)\equiv x^3-x^2+2x+7=0.
\ee
One readily gets the roots $x_a$, $x_b$, $x_c$, of the polynom $g$, so that
\be
g(x)\equiv (x-x_a)(x-x_b)(x-x_c),
\ee
where
\be
\label{eqtfo}
\begin{array}{l}
x_a=\frac{1}{3}+Y_1+Y_2,\\[+0.1cm]
x_b=\frac{1}{3}+\epsilon Y_1+\epsilon^2 Y_2,\\[+0.1cm]
x_c=\frac{1}{3}+\epsilon^2 Y_1+\epsilon Y_2,
\end{array}
\ee
with {\em real} quantities 
\be
\label{eqtfi}
Y_{1,2}=\pm\frac{(20)^\frac{1}{3}}{6}\left(\mp 41+9\sqrt{21}\right)^{\frac{1}{3}},
\ee
and the complex third root of unity
\be
\label{eqtsi}
\epsilon=e^{i\pi /3}=-\frac{1}{2}+i\frac{\sqrt{3}}{2}.
\ee
It is worth to observe that $x_a$ is the only real root of $g$, since $g(x)$ is an increasing function because its derivative is positive. Also, $g(-2)<0$ and $g(-1)>0$, so that the root $x_a\in(-2,-1)$. The other two roots are mutually conjugated, i.e. $x_b=x_c^{*}$.

Now we note that the formula (\ref{eqto}) associates two roots
\be
\label{eqtse}
t_{1,2}^x=\frac{x}{2}\pm\frac{i}{2}\sqrt{4-x^2}, \quad x\in\{x_a,x_b,x_c\},
\ee
of the polynom $f$ with each root $x$ of the polynom $g$, by means of the corresponding new polynom
\be
\label{eqtsei}
\phi^x(t)=t^2-xt+1\equiv (t-t^x_1)(t-t^x_2), \quad x\in\{x_a,x_b,x_c\}.
\ee
One thus has the factorization 
\be
\label{eqtni}
f(t)=\phi^{x_a}(t)\phi^{x_b}(t)\phi^{x_c}(t),
\ee
with the corresponding decomposition of the set $R^f$ of all roots of the polynom $f$ into those of polynoms $\phi^x$
\be
\label{eqthirty}
R^f=\{a_1,a_2\}\cup \{b_1,b_2\}\cup \{c_1,c_2\}.
\ee
All these portions of phase in $R^f$ are given explicitely by Eqs. (\ref{eqtfo})-(\ref{eqtse}).

We point out some simple algebraic symmetries of the set $R^f$ of all relevant portions of phase. First, it follows from Eq. (\ref{eqto}) that
\be
\label{eqthirtyo}
t_2^x=(t_1^x)^{-1},
\ee
so that each set in rhs of Eq. (\ref{eqthirty}) consists of mutual inverses ($a_2=a_1^{-1}$, etc.). Next, the reality of $x_a$ and $|x_a|<2$ implies 
\be
\label{eqthirtytwo}
|a_1|=|a_2|=1,
\ee
or by virtue of Eq. (\ref{eqtw}), the real pseudomomenta, whereas the fact that $x_b=x_c^{*}$ yields that $\{c_1,c_2\}$ are complex conjugates of $\{b_1,b_2\}$. Taken together, the only solutions of the inverse BA for the qubit $\mathcal{H}^{33}_0$ are either
\be
\label{eqthirtythree}
(a,b,c)=\left(\frac{b_1^{*}}{b_1},b_1,\frac{1}{b_1^{*}}\right),
\ee
or
\be
\label{eqthirtyfour}
(a,b,c)=\left(\frac{b_1}{b_1^{*}},\frac{1}{b_1},b_1^{*}\right),
\ee
with $b_1=t_1^{x_b}$.

We have thus obtained exact results for values of six Bethe parameters, represented by portions of phase, which fully determine the two Bethe eigenstates within the qubit $\httz$. So, we are in a good position to make a comparison with the hypothesis of strings. According to combinatoric prescription \cite{llls}, the string configurations rigged by quasimomenta are
\be
\label{eqthirtyfiven}
\begin{array}{c}
\raisebox{2pt}{$\nu \mathcal{L} \equiv$}
\\
\end{array}
\begin{array}{c}
\begin{Young}
$L_2$  &  \cr
$L_1$     \cr
\end{Young}\\
\end{array}
=
\left\{
\begin{array}{l}
\begin{Young}
$3$  &  \cr
$-3$     \cr
\end{Young}\\
\begin{Young}
$-3$  &  \cr
$3$     \cr
\end{Young}\\
\end{array}
\begin{array}{l}
\mbox{for\,}v=1,\\[+0.5cm]
\mbox{for\,}v=2,
\end{array}
\right.
\ee
where $v$ are labels of the two BA eigenstates. In other words, these two eigenstates belong to the string configuration $\nu=\{21\}$, that is, a 2-string and a 1-string, and each string is rigged by the quasimomentum $L=\pm 3\in B$, such that the total quasimomentum is zero. Clearly, Eqs. (\ref{eqthirtythree}) and (\ref{eqthirtyfour}) fully confirm the quantum number $\nu$ of string configuration: Eq. (\ref{eqthirtytwo}) indicates the one-string, whereas
\be
\label{eqthirtysixn}
|b_1c_1|=|b_2c_2|=1
\ee
points out the two-string. We can thus make the following assignment of portions of phase to the boxes of the Young diagram $\nu\vdash 3$

\be
\label{eqkratkiok}
\begin{array}{|c|c|}
\hline
b  & c \\
\hline
a     \\
\cline{1-1}
\end{array}
\quad = \quad \mbox{either} \quad 
\begin{array}{|c|c|}
\hline
b_1  & 1/b_1^{*} \\
\hline
b_1^{*}/b_1     \\
\cline{1-1}
\end{array}\, , 
\quad \mbox{or} \quad 
\begin{array}{|c|c|}
\hline
1/b_1  & b_1^{*} \\
\hline
b_1/b_1^{*}     \\
\cline{1-1}
\end{array}\, .
\ee

\noindent We have therefore $a_{1,2}=t^{x_a}_{2,1}$, $b_{1,2}=t^{x_b}_{1,2}$, and $c_{1,2}=t^{x_c}_{1,2}$, in accordance with requirements of the inverse BA.\\

The set $R^f$ of all roots of the polynom $f$, and therefore of all relevant portions of phase, is depicted in Fig. 2
. Eq. (\ref{eqtni}) presents a factorization of $f$ along the constituents of strings, represented by boxes of the string configuration $\nu$ in Eq. (\ref{eqkratkiok}). This factorization displays the vertical fibration of $R^f$ in Fig. 2.
\begin{figure}[ht]
\begin{center}
\includegraphics*[width=0.25\linewidth]{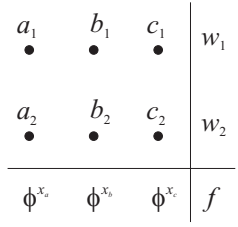}
\label{fig_polynf}
\end{center}
\caption{Presentation of the set $R^f$ of all roots of the polynom $f$. It displays admissible Bethe parameters (portions of phase) for the qubit $\mathcal{H}^{33}_0$. The polynom $f$ can be presented as the product of either three polynoms $\phi^{x_a}$, $\phi^{x_b}$, $\phi^{x_c}$ (the vertical fibration, representing the constituents of strings), or two polynoms $w_1$, $w_2$ (the horizontal fibration, which separates the two BA eigenstates in the qubit entering each of the two BA states).}
\end{figure}

There is another important factorization, given by
\be
\label{eqthirtysix}
f(t)=w_1(t)w_2(t),
\ee
where 
\be
\label{eqthirtyseven}
\begin{array}{l}
w_1(t)=(t-a_1)(t-b_1)(t-c_1)\equiv t^3-\left(\frac{1}{2}+\frac{i}{2}\sqrt{15}\right)t^2+\left(\frac{1}{2}-\frac{i}{2}\sqrt{15}\right)t-1\, ,\\[+0.1cm]
w_2(t)=(t-a_2)(t-b_2)(t-c_2)\equiv t^3-\left(\frac{1}{2}-\frac{i}{2}\sqrt{15}\right)t^2+\left(\frac{1}{2}+\frac{i}{2}\sqrt{15}\right)t-1\, .
\end{array}
\ee
It corresponds to the horizontal fibration of $R^f$. Three roots of each polynom, $w_1$ and $w_2$, define a unique BA state in the qubit $\mathcal{H}^{33}_0$. We conclude that there are exactly two such BA states within this qubit.

Once the portions of phase are known, the corresponding spectral parameters are readily derived from the Cayley transform (\ref{eqjacobi}). They can be presented \linebreak in a compact form as 
\be
\label{eqthirtyeight}
\lambda_{1,2}^x=\pm\frac{1}{\sqrt{15}}\left(\frac{1}{2}+x\right), \quad x\in\left\{x_a,x_b,x_c\right\},
\ee
where the index 1 (2) corresponds to the upper (lower) sign and the first (second) BA eigenstate. 

It is convenient to define polynomials $u_1$ and $u_2$ as
\be
\label{eqthirtynine}
\begin{array}{l}
u_1(\lambda)=(\lambda-\lambda_1^a)(\lambda-\lambda_1^b)(\lambda-\lambda_1^c)\equiv \lambda^3-\frac{5}{2\sqrt{15}}\lambda^2+\frac{1}{4}\lambda+\frac{3}{8\sqrt{5}}\, ,\\[+0.1cm]
u_2(\lambda)=(\lambda-\lambda_2^a)(\lambda-\lambda_2^b)(\lambda-\lambda_2^c)\equiv \lambda^3+\frac{5}{2\sqrt{15}}\lambda^2+\frac{1}{4}\lambda-\frac{3}{8\sqrt{5}}\, ,
\end{array}
\ee
with upper indices $x_a,x_b,x_c$ replaced for simplicity by $a,b,c$, respectively. We use them in the next chapter to reduce considerably degrees of polynomials of spectral parameters in highly nonlinear calculations of algebraic BA.

Spectral parameters provide another check of the string hypothesis. For further purpose, we present these parameters in a more transparent form, as $(\lambda_{1,2}^a,\lambda_{1,2}^b,\lambda_{1,2}^c)\equiv(\lambda_{1,2},\mu_{1,2},\nu_{1,2})$. It follows that they can be written in the form
\be
\label{eqfourtytwon}
\begin{array}{c}
\lambda_1=-\lambda_2=\lambda_0,\\
\mu_1=-\mu_2=\mu_0+im,\\
\nu_1=-\nu_2=\mu_0-im,
\end{array}
\ee
with $\lambda_0$, $\mu_0$, and $m$ real. Thus, in terms of spectral parameters, one has
\be
\label{eqfourtythreen}
\begin{array}{|c|c|}
\hline
\mu  & \nu \\
\hline
\lambda     \\
\cline{1-1}
\end{array}
\quad = \quad \mbox{either} \quad 
\begin{array}{|c|c|}
\hline
\mu_0+im  & \mu_0-im \\
\hline
\lambda_0     \\
\cline{1-1}
\end{array}\, , 
\quad \mbox{or} \quad 
\begin{array}{|c|c|}
\hline
-\mu_0-im  & -\mu_0+im \\
\hline
-\lambda_0     \\
\cline{1-1}
\end{array}\, .
\ee
It is the form predicted literally by the string hypothesis, with the only exception that
\be
\label{eqfourtyfourn}
m=\frac{1}{2\sqrt{5}}(Y_1-Y_2)\approx 0,503,
\ee
which differs slightly from the asymptotic value $m=1/2$ for the thermodynamic limit \cite{faddeevrus}. 

In order to evaluate riggings $L_1$ and $L_2$ of the 1-string and 2-string of Eq. (\ref{eqthirtyfiven}), and thus to complete the inverse BA, we use BA equations in the form (\ref{eqn}). We rewrite them as 
\be
\label{eqfourtyfivenabc}
\begin{array}{c}
a^7=V(a,b)V(a,c),\\
b^7=V(b,a)V(b,c),\\
c^7=V(c,a)V(c,b),
\end{array}
\ee
with 
\be
\label{eqfourtysixn}
V(a,b)=\frac{ab-2a+1}{ab-2b+1}=\frac{1}{V(b,a)}
\ee
describing the scattering the of pseudoparticle associated with $a$ on that for $b$. By virtue of construction, these equations are exactly satisfied by portions of phase (\ref{eqthirtythree}) and (\ref{eqthirtyfour}). In order to determine pseudomomenta nad riggings, one has to evaluate the logarithms of Eq. (\ref{eqfourtyfivenabc}). To this aim we put
\be
\label{eqfourtysevenn}
b_1=be^{i\beta}\equiv e^{i(p'+ip'')},
\ee
with real amplitude $b=e^{-p''}$ and phase $\beta=p'$, so that the complex pseudomomentum associated with the first box of the 2-string in $\nu$ is $p'+ip''$. Then one has 
\be
\label{eqfourtyeightn}
(a,b,c)=\mbox{\,either\,} (e^{-2i\beta},be^{i\beta},b^{-1}e^{i\beta}), \mbox{\,or\,} (e^{2i\beta},b^{-1}e^{-i\beta},be^{-i\beta}),
\ee
so that the pseudomomentum of the 1-string is $\pm 2\beta$, and the total pseudomomentum of the 2-string is $\mp 2\beta$. The rigging $L_1$ of the 1-string is determined from the logarithm of Eq. (\ref{eqfourtyfivenabc}) as 
\be
\label{eqfourtyninen}
7\cdot(\pm 2\beta)-\phi_{1,2}=2\pi L_1,
\ee
where $\pm 2\beta$ is the pseudomomentum of the 1-string, and $\phi_{1,2}$ is the phase of scattering, given by 
\be
\label{eqfiftyn}
V(a_{1,2},b_{1,2})V(a_{1,2},c_{1,2})=e^{i\phi_{1,2}}
\ee
(observe that $|V(a_{1,2},b_{1,2})V(a_{1,2},c_{1,2})|=1$).\\
\noindent The rigging $L_2$ of the 2-string follows from multiplication of both sides of Eq. (\ref{eqfourtyfivenabc}), i.e.
\be
\label{eqfiftyonen} 
b^7c^7=V(b,a)V(c,a)=e^{-i\phi_{1,2}}.
\ee
The corresponding logarithm yields
\be
\label{eqfiftytwon}
7\cdot(\mp 2\beta)+\phi_{1,2}=2\pi L_2.
\ee
Note that $\beta$ and $\phi_{1,2}$, as well as other Bethe parameters, have attached exact values. We do not quote them due to their curiosity, but present the resulting riggings as
\be
\label{fiftythreen}
(L_1,L_2)=\mbox{\,either\,} (3,-3), \mbox{\,or\,} (-3,3),
\ee
so that the 1-string and 2-string have opposite riggings, maximal within Brillouin zone for heptagon.

\section{Algebraic Bethe Ansatz}
We proceed to derive an explicit form of BA eigenstates for the qubit $\mathcal{H}^{33}_0$, using techniques of algebraic BA \cite{faddeevrus} - \cite{fadd}. The main tool there is the monodromy matrix $\mathcal{M}(\lambda)$, which is a $\lambda$-dependent operator acting in the space $\mathcal{H}\otimes V$, with $V\cong \mathbb{C}^2$ usually referred to as ''the auxiliary space''. It is defined as the product
\be
\label{eqfourty}
\mathcal{M}(\lambda)=L_7(\lambda)L_6(\lambda)\ldots L_1(\lambda)
\ee
of Lax operators $L_j(\lambda)$ along the heptagon ($j\in\tilde{7}$). The Lax operator, written as a matrix in the auxiliary space $V$, has the form
\be
\label{eqfourtyone}
L_j(\lambda)=
\left(
\begin{array}{cc}
a_j(\lambda)&b_j\\
c_j&d_j(\lambda)
\end{array}
\right)\, ,
\ee
where 
\be
\label{eqfourtytwo}
a_j(\lambda)=I\lambda+\frac{i}{2}s^z_j, \quad b_j=is_j^{-}, \quad c_j=is_j^{+}, \quad d_j(\lambda)=I\lambda-\frac{i}{2}s^z_j,
\ee
are operators in $\mathcal{H}$, with $s_j^z$, $s_j^{\pm}$ being the spin operators for the node $j\in\tilde{7}$, and $I$ - the identity operator in $\mathcal{H}$. Nowadays, it is a simple calculational matter to evaluate explicitely the monodromy matrix in a computer, and to present the result in the form
\be
\label{eqfourtythree}
\mathcal{M}(\lambda)=
\left(
\begin{array}{cc}
A(\lambda)&B(\lambda)\\
C(\lambda)&D(\lambda)
\end{array}
\right)\, ,
\ee
where $A(\lambda),B(\lambda),C(\lambda),D(\lambda)$ are explicitely known operator-valued functions of the spectral paremeter $\lambda$. Clearly, the Lax operators (\ref{eqfourtytwo}) are local, i.e. they act effectively only in the space $\left(\mathbb{C}^2\right)_j$ for the $j$-th node, whereas elements $A,B,C,D$ of the monodromy matrix (\ref{eqfourtythree}) are global. In the following, we exploit the property that the operator $B(\lambda)$, when acting on the vacuum state \linebreak $|+\ldots+\rangle=|0\rangle$, creates the one-deviation state characterized by the spectral parameter $\lambda$, or the corresponding pseudomomentum $p$ (cf. Eq. (\ref{eqf})). More generally, $B(\lambda)B(\mu)B(\nu)|0\rangle$ is an (unnormalized) state in the sector $\mathcal{H}^3$, characterized by the collection $\{\lambda,\mu,\nu\}$ of spectral parameters (the latter should be distinct pairwise). We use this method for construction of BA eigenstates for the qubit $\mathcal{H}^{33}_0$, specified by collections of spectral parameters, determined in the previous section.

A three-magnon state, characterized by arbitrary values of the collection $\{\lambda,\mu,\nu\}$ of spectral parameters (finite, $\mathbb{C}$-valued, pairwise distinct, different from $\pm i/2$) can be written as
\be
\label{eqffnab}
\begin{array}{rcl}
|\{\lambda,\mu,\nu\}\rangle&=&B(\lambda)B(\mu)B(\nu)|0\rangle\\[+0.1cm]
&=&B^{32}(\lambda)B^{21}(\mu)B^{10}(\nu)|0\rangle,
\end{array}
\ee
where $B(\lambda)$, $B(\mu)$, $B(\nu)$ in (\ref{eqffnab}) are operators in the whole space $\mathcal{H}$, whereas $B^{r,r-1}$, $r=1,2,3$ in (\ref{eqffnab}) are rectangular blocks of the size $\left( {N \atop r} \right) \times \left( {N \atop r-1} \right)$, representing the corresponding homomorphisms from $\mathcal{H}^{r-1}$ to $\mathcal{H}^{r}$ (all the other blocks of $B(\lambda)$ are either zeros, or outside the equator, and thus irrelevant for our purposes). 
In accordance with definitions (\ref{eqfourty}) - (\ref{eqfourtythree}), a matrix element of the block $B^{r,r-1}$ is either zero, or a monomial of degree $N$ in three complex numbers, $p$, $q$, and $i$, where
\be
\label{eqfourtyfive}
p=\lambda+\frac{i}{2}, \quad q=\lambda-\frac{i}{2}
\ee
are eigenvalues of diagonal elements of the (operator valued) Lax matrix (\ref{eqfourtyone}) in the basis of magnetic configurations of the heptagon, whereas the factor $i$ emerges from its non-diagonal elements. The well known fact of commutativity,
\be
\label{eqfourtysix}
[B(\lambda),B(\mu)]=0
\ee
implies that the state $|\{\lambda,\mu,\nu\}\rangle$ does not depend on the ordering of spectral parameters. Thus this state is a symmetric function of these parameters, even if the way of construction (\ref{eqffnab}) clearly depends upon the order.

In particular, the block $B^{10}(\lambda)$ has the form
\be
\label{eqfourtyseven}
B^{10}(\lambda)=\left(
\begin{array}{c}
ip^6\\
ip^5q\\
ip^4q^2\\
ip^3q^3\\
ip^2q^4\\
ipq^5\\
iq^6
\end{array}
\right),
\ee
which yields the state
\begin{eqnarray}
\label{eqfourtyen}
|\{\lambda\}\rangle=B(\lambda)|0\rangle&=\sum_{j\in\tilde{7}}ip^{7-j}q^{j-1}|j\rangle=\\[+0.1cm]
\label{eqfourtyenn}
&=ip^7q^{-1}\sum_{j\in\tilde{7}}a^{-j}|j\rangle,
\end{eqnarray}
where $|j\rangle\in\mathcal{H}^{1}$ is the magnetic configuration with the single spin deviation at the node $j\in\tilde{7}$. $|\{\lambda\}\rangle$ is readily recognized as the one-magnon (unnormalized) state characterized by the portion of phase
\be
\label{eqfifty}
a=\frac{p}{q}=\frac{\lambda+\frac{i}{2}}{\lambda-\frac{i}{2}}.
\ee
Clearly, for $a=\mbox{exp\,}(2\pi ik/7)$, $k\in B$, it is the eigenstate of the Heisenberg Hamiltonian (\ref{eqtw}), corresponding to magnon with the quasimomentum $k$. 

Such a transparent and easy calculation of the block $B^{10}(\lambda)$ within algebraic BA formalism sheds some light how to overcome calculations aimed at exact results for three spin deviations. It is clear from Eq. (\ref{eqfourtyseven}) that each matrix element of the block $B^{10}(\lambda)$ is a monomial of degree 6 with respect to spectral parameter $\lambda$, Moreover, it readily follows from the general rules (\ref{eqfourty}) - (\ref{eqfourtythree}) that (i) the creation operator  $B(\lambda)$ is a sum of $N$-th degree monomials of the Lax objects (\ref{eqfourtytwo}), one object for each node, (ii) the only non-vanishing terms in the expression $B(\lambda)|0\rangle$ have the form
\be
\label{eqfiftyone}
d_1\ldots d_{j-1}b_ja_{j+1}\ldots a_N|0\rangle=ip^{N-j}q^{j-1}|j\rangle,
\ee
with a single creation operator $b_j$ of the spin deviation at the node $j$, preceded by $j-1$ diagonal Lax objects $d_{j'}$, $j'=1,\ldots,j-1$, and followed by $N-1$ objects $a_{j'}$, $j=j+1,\ldots,N$. It readily yields the result (\ref{eqfourtyen}) - (\ref{eqfourtyenn}). Essentially similar considerations provide the form of blocks $B^{21}(\lambda)$ and $B^{32}(\lambda)$, with somehow increased combinatoric complexity emerging from the fact that now the Lax objects (\ref{eqfourtytwo}) do not act on the ferromagnetic vacuum $|0\rangle$. But the whole derivation, based on Eqs. (\ref{eqfourty}) - (\ref{eqfourtythree}), can be easily and precisely done on a computer. Matrix elements of these blocks are either zeros, or monomials in $\lambda$ of degree $6$, $4$, or $2$.

Now we are in a position to use the results of Section 3 for further simplification of these blocks. Namely, we have pointed out that the spectral parameters of each of the two exact eigenstates of the qubit $\httz$ satisfy a polynomial equation of degree 3, given explicitely by Eq. (\ref{eqthirtynine}). Thus all powers $\lambda^l$, $l\geq 3$ can be expressed uniquely in terms of $\lambda^0=1$, $\lambda$, and $\lambda^2$. For example, 
\be
\label{eqfiftytwo}
p^3q^3=\lambda^6+\frac{3}{4}\lambda^4+\frac{3}{16}\lambda^2+\frac{1}{64}\cong
\left(\frac{17}{72}-\frac{\sqrt{3}}{8}\right)\lambda^2 \mp \left(\frac{\sqrt{5}}{20}-\frac{\sqrt{15}}{36}\right)\lambda + \frac{7}{160}-\frac{\sqrt{3}}{24},
\ee
where the upper and lower sign corresponds to residuum modulo $u_1$ and $u_2$ of Eq. (\ref{eqthirtynine}), respectively. In this way, we reach the matrix elements of each block as polynoms of degree at most 2, with considerably reduced the annoying nonlinearity of BA formalism, while keeping the results exact.

Using Eq. (\ref{eqffnab}), we obtain an unnormalized 35-component vector \linebreak $|\{\lambda_1,\lambda_2,\lambda_3\}\rangle\in\mathcal{H}^3$, whose elements (in the basis of $C_7$ orbits $|\bft j\rangle$, $\bft \in V,j\in\tilde{7}$) are symmetric functions of $\{\lambda,\mu,\nu\}$, with the degree of each not exceeding 2. By applying the Fourier transform $F^{V3}:\mathcal{H}^3\to\mathcal{H}^3_0$, in a form of rectangular $5\times 35$ matrix with elements 
\be
\label{eqfiftythree}
F^{V3}_{\bft,\bft'j}=\delta_{\bft,\bft'}, \quad \bft, \bft'\in V, j\in\tilde{7}
\ee
(the transform $F^{V3}$ is also unnormalized), we obtain a 5-component vector $F^{V3}|\{\lambda,\mu,\nu\}\rangle\in\mathcal{H}^3_0$. Then, substituting appropriate numerical values of spectral parameters (\ref{eqthirtyeight}) in accordance with (\ref{eqthirtythree}) - (\ref{eqthirtyfiven}), we obtain the desired BA eigenstates. Let $\rho_1$ and $\rho_2$ be the density matrices, corresponding to the sets (\ref{eqthirtythree}) and (\ref{eqthirtyfour}) of Bethe parameters. They are given by
\be
\label{eqfiftyfour}
\rho_1=
\left(
\begin{array}{cccccc}
A&A&A&\shortmid&B&B^{*}\\
A&A&A&\shortmid&B&B^{*}\\
A&A&A&\shortmid&B&B^{*}\\
\mbox{--}&\mbox{--}&\mbox{--}&\mbox{+}&\mbox{--}&\mbox{--}\\
B^{*}&B^{*}&B^{*}&\shortmid&6A&C\\
B&B&B&\shortmid&C^{*}&6A\\
\end{array}
\right), \quad \rho_2=\rho_1^{*},
\ee
with 
\be
\label{eqfiftyfive}
A=\frac{2}{30}, \quad B=\frac{-3+i\sqrt{5}}{30}, \quad C=\frac{-1+i\sqrt{15}}{10},
\ee
and the asterisk $*$ denoting the complex conjugation. 
It is worth to observe that the exact result (\ref{eqfiftyfour}) was otained by use of algebraic BA, together with a combinatoric analysis of roots of the polynom $f$ (Eq. (\ref{eqtz})), derived from "inverse BA". A good check of calculations is provided by the sum rule
\be
\label{eqfiftysix}
\rho_1+\rho_2=P_0^{33},
\ee
where $P_0^{33}$ is the projector onto the qubit $\httz$, given by Eq. (\ref{eqeightt}). Also,
\be
\label{eqfiftyseven}
\rho_1\rho_2=0,
\ee
so that BA eigenstates are mutually orthogonal.\\

The result (\ref{eqfiftyfour}) justifies a postiriori the subdivision (\ref{eqnine}) of the set $V$ of wavelets into subsets $V_1$ and $V_2$. A comparison with Eq. (\ref{eqeightt}) and Fig. \ref{fig_basisk} points out that each element of each of these subsets enters the density matrices of Eq. (\ref{eqfiftyfour}) on equal footing: with the same occupation numbers ($A$ and $6A$ for $V_1$ and $V_2$, respectively), and the same hybridization parameters (real $A$, complex $C$, and complex $B$ for internal hybridization within subset $V_1$, the same for $V_2$, respectively). It is worth to observe that the elements of the subset $V_1$ differ mutually by kinematics (distinct structures of islands of adjacent spin deviations) and dynamics (distinct structures of interaction channels - cf. Fig. \ref{fig_basisk}). Their common feature is invariance with respect to the parity $\pi$ (cf. Eqs. (\ref{eqten}) and (\ref{eqeleven})), whereas the subset $V_2$ consists of two enantiomorphic elements.

Total occupation of the subset $V_1$ is $3A=1/5$, whereas that of $V_2$ is $12A=4/5$. The internal hybridization within the subset $V_1$ is also given by the real parameter $A$, so that there is no net probability current within this subset. Such a current exists both between the two elements of the subset $V_2$, as well as between $V_1$ and $V_2$, owing to the complex values of $B$ and $C$. Clearly, currents corresponding to both BA eigenstates $\rho_1$ and $\rho_2$ have opposite signs.\\

We mention at the end that the arithmetic qubit considered here provides the simplest demonstration of the fact that strings of different length are {\em distinguishable} objects, i.e. the interchange if the 1-string with the 2-string produces a quantum state which is distinct from the initial one (cf. Eq. (\ref{eqthirtyfiven})). It can be contrasted with the two-magnon sector of the heptagon, where one encounters the rigged string configurations
\be
\label{eqseventytwon}
\left|
\begin{array}{c}
\begin{Young}
$2$   \cr
$-2$     \cr
\end{Young}\\
\end{array}
\right\rangle
\quad \mbox{and} \quad 
\left|
\begin{array}{c}
\begin{Young}
$3$   \cr
$-3$     \cr
\end{Young}\\
\end{array}
\right\rangle
\ee
Now, the interchange of the two 1-strings in each of these states does not produce any distinct state: each of these states is invariant under parity or is selfenantiomorphic. It demonstrates that the 1-strings are indistinguishable entities.

\section{Conclusions}
We have examined the BA form of exact eigenfunctions of the heptagon within the XXX model for a specific case when an extra symmetry (outside spherical and translational) admits a degenarate eigenspace of Heisenberg Hamiltonian, with the same energy, quasimomentum and the total spin. This eigen-space realizes an arithmetic qubit at the centre of the Brillouin zone, such that each element of this qubit realizes a legitimate exact eigenstate of the Heisenberg Hamiltonian, but only some of them have the form prescribed by BA. Indeed, as we have shown, only two states of this arithmetic qubit have the desired BA form, and can thus be presented in terms of rigged string configurations. 

We have determined explicitely Bethe parameters of the two BA eigenstates, using the so called "inverse BA", and derived the corresponding density matrices using algebraic BA. We have shown that the degeneracy of this arithmetic qubit has its origin in invariance with respect to the parity symmetry of the heptagonal ring. This invariance has a clear presentation within the picture of rigged string configurations: the action of the parity operator results there in the change of sign of rigging (i.e. quasimomentum) of each constituent string, and the exchange of the 2-string with the 1-string of one of BA eigenstates produces the second eigenstate. 



\newpage

\begin{thebibliography}{99}
\bibitem{bethe} H. Bethe, Z. Physik 71 (1931) 205 (in German; English translation in: D.C. Mattis, The Many-Body Problem, Singapore, World Sci., 1993, pp. 689-716).
\bibitem{takahashi} M. Takahashi, Progr. Theor. Phys. 46 (1971) 401.
\bibitem{yangone} C.N. Yang, C.P. Yang, Phys. Rev. 150 (1966) 321.
\bibitem{vladimirov} A.A. Vladimirov, Phys. Lett. 105 A (1984) 418.
\bibitem{essler} F.H.L. Essler, V.E. Korepin, K. Schoutens, J. Phys. A 25 (1992) 4115. 
\bibitem{juttner}
G. J\"uttner, B.-D. D\"orfel, J. Phys. A: Math. Gen. 26 (1993) 3105.
\bibitem{warnaar} S.O. Warnaar, J. Stat. Phys. 82 (1996) 657.
\bibitem{baxter1} R.J. Baxter, J. Stat. Phys. 108 (2002) 1. 
\bibitem{clwkl} W.J. Caspers, M. {\L}abuz, A. Wal, M. Ku\'zma, T. Lulek, J. Phys. A: Math. Gen. 36 (2003) 5369.
\bibitem{hage} B. Hagemans, J.-S. Caux, J. Phys. A: Math. Gen. 40 (2007) 14605.
\bibitem{kkr} S.V. Kerov, A.N. Kirillov, N.Yu. Reshetikhin, LOMI 155 (1986) 50
(in Russian; English translation: J. Sov. Math. 41 (1988) 916). 
\bibitem{llls} B. Lulek, T. Lulek, M. Labuz, R. Stagraczynski, Physica B 405 (2010) 2654.
\bibitem{mbll} J. Milewski, G. Banaszak, T. Lulek, M. Labuz, Physica B (2011) 520.
\bibitem{mblls} J. Milewski, G. Banaszak, T. Lulek, M. \L{}abuz, R. Stagraczy\'nski, OSID 19 (2012) 1250012.
\bibitem{llwj} B. Lulek, T. Lulek, A. Wal, P. Jakubczyk, Physica B 337 (2003) 375.
\bibitem{faddeevrus} L.D. Faddeev, L.A. Takhtajan, LOMI 109 (1981) 134
(in Russian; English translation: J. Sov. Math. (1984) 241).
\bibitem{fadd} L.D. Faddeev, arXiv:hep-th/9605187v1.
\end{thebibliography}
\end{document}